\documentclass[twocolumn,showpacs,preprintnumbers,amsmath,amssymb,aps]{revtex4}
\usepackage{mathrsfs}
\usepackage[hypertex,linkcolor=red]{hyperref}
\def\d{\operatorname{d}}\def\<{\langle}\def\>{\rangle}
\def\Tr{\operatorname{Tr}}\def\Span{\operatorname{Span}}\def\Supp{\operatorname{Supp}}
\def\sH{\set{H}}\def\sW{\set{W}}\def\Base{\set{B}}
\def\Lik{\mathcal{L}}\def\grp#1{{\mathbf #1}}\def\gG{\grp{G}}
\def\map#1{{\mathscr{#1}}}\def\set#1{{\sf #1}}\def\param{x}\def\estparam{\hat{x}}
\def\estg{\hat{g}}
\def\Klm{{\mathfrak X}}\def\Ism{I}\def\Id{I}
\def\dim{\operatorname{dim}}
\def\Cmplx{\mathbb C}\def\System{\mathcal{S}}\def\matrix{\mathbb{M}}
\begin{document}
\pacs{03.65.Ta 03.67.-a 02.20.-a} \title{Covariant quantum
  measurements which maximize the likelihood} \author{Giulio
  Chiribella} \email{chiribella@fisicavolta.unipv.it} \author{Giacomo
  Mauro D'Ariano} \email{dariano@unipv.it} \altaffiliation[Also at
]{Center for Photonic Communication and Computing, Department of
  Electrical and Computer Engineering, Northwestern University,
  Evanston, IL 60208} \author{Paolo Perinotti}
\email{perinotti@fisicavolta.unipv.it} \author{Massimiliano F. Sacchi}
\email{msacchi@unipv.it} \affiliation{QUIT (Quantum Information
  Theory) Group of the INFM, unit\`a di Pavia}
\homepage{http://www.qubit.it} \affiliation{Dipartimento di Fisica
  ``A. Volta'', via Bassi 6, I-27100 Pavia, Italy} \date{\today}
\begin{abstract}
  We derive the class of covariant measurements which are optimal
  according to the maximum likelihood criterion. The optimization
  problem is fully resolved in the case of pure input states, under
  the physically meaningful hypotheses of unimodularity of the
  covariance group and measurability of the stability subgroup.  The
  general result is applied to the case of covariant state estimation
  for finite dimension, and to the Weyl-Heisenberg displacement
  estimation in infinite dimension. We also consider estimation with
  multiple copies, and compare collective measurements on identical
  copies with the scheme of independent measurements on each copy. A
  "continuous-variables" analogue of the measurement of direction of
  the angular momentum with two anti-parallel spins by Gisin and
  Popescu is given.
\end{abstract}
\maketitle
\section{Introduction}
State estimation is a unique kind of quantum measurement in the
quality of information that it provides. In fact, the knowledge of the
state of a quantum system enables the evaluation of any ensemble
average, which is equivalent to the possibility of performing any
desired experiment on the system. For its intrinsic versatility such
unconventional type of quantum measurement is of interest for the new
technology of quantum information \cite{Nielsen2000} in the estimation
of parameters that do not correspond to
observables \cite{Helstrom76}---such as the phase of an electromagnetic
field  \cite{buggec}---but also as a method to achieve quantum
cloning \cite{bem,joint_sub}, whence in designing eavesdropping
strategies for quantum cryptography \cite{gisinrev}.
\par An exact state estimation without any prior knowledge 
on the form of the state is impossible \cite{single} due to the
no-cloning theorem \cite{WoottersZurek,Yuen}. This also reflects the
fact that an optimal approximate state estimation would not be
achievable as an orthogonal measurement, since the state estimation is
a kind of "informationally complete" measurement \cite{univest}. More
generally, one can have some prior knowledge of the form of the state,
i.e. by parameterizing it with a restricted set of variables. This is
the typical situation of the Quantum Estimation Theory of
Helstrom \cite{Helstrom76}, where the goal is to determine a
multidimensional parameter of a state transformation. When the set of
states to be discriminated are orthonormal the parameter corresponds
to an "observable" whose eigenstates are the set itself, and the
estimation is exact. However, in practice it happens very often that
the multidimensional parameter cannot be described by an observable
(e.g. it is a phase of a field, or it corresponds to a set of non
compatible observables), whence a measurement represented by a
so-called {\em positive operator valued measure} (POVM) needs to be
performed.
\par For a state estimation which is not equivalent to the measurement of an
observable we have a choice of infinitely many POVM's achieving the
same task with different strategies. Indeed there is no universal
criterion which is optimal for all situations, and one needs to define
the appropriate figure of merit pertaining to the particular problem.
Once the optimization problem is solved in terms of an optimal POVM,
one can then address the problem of the feasibility of the measurement
apparatus by classification of orthogonal dilations of the POVM
 \cite{Helstrom76,Nai,busc}, or else compare the performance of actual
devices to the ultimate theoretical limit.
\par A statistically meaningful optimization strategy is the maximization of
the likelihood that the true value of the estimated parameter
coincides with the outcome of the measurement. Such a strategy is
actually very general, since for measurements which are {\em
  group-covariant}, optimization of a generic {\em goal function}
corresponds to optimization of the likelihood for a different input
state.  Physically, "group-covariance" means that there is a group of
transformations on the probability space which maps events into
events, in such a way that when the quantum system is transformed
according to one element of the group, the probability of the given
event becomes the probability of the transformed event. This situation
is very natural, and occurs in most practical applications. For
example, the heterodyne measurement \cite{YuenSha,Bilk-poms} is
covariant under the group of displacements of the complex field, which
means that if we displace the state of radiation by an additional
complex averaged field, then the output photo-current will be
displaced by the same complex quantity. Other examples of covariant
measurements are the quantum estimation of a "spin orientation"
\cite{direzionespin,peres,tapia1,tapia2}, or of the phase shift of an
electromagnetic field \cite{Helstrom76,sasschind,olevo}.
\par The statistics of the measurement can be improved by using many
copies of the same quantum system. In this scenario, it is relevant
for experiments to distinguish the measurements achievable by local
operations and classical communication (LOCC) from more general
schemes that require entanglement. Unfortunately, a useful
classification of LOCC schemes is still missing. Alternatively, one
can give just a mathematical categorization in terms of the POVM of
the measurement: i) ``independent'' measurements, corresponding to
tensor product of independent POVM's; ii) ``separable'' measurements,
corresponding to POVM's where each element is separable; iii) ``non
separable'' or ``entangled'' measurements, corresponding to POVM's
where some element is entangled. In the first category measurements
are performed independently on each copy.  In the separable class, on
the other hand, the measurement can be performed by means of separable
operations, hence all LOCC schemes are included in this category.
Notice, however, that not all separable operations can be implemented
locally (see, e.g., the case of nonlocality without entanglement of
Ref.  \cite{noent}). Finally, the class of entangled POVM's represents
the most general scheme of measurement, and opens the exponential
growth of the Hilbert space dimension versus the number of copies $N$,
with the possibility of largely surpassing the statistical efficiency
of the independent measurement schemes
 \cite{buggec,Chefles,lopar,olevoasymptotic}. However, as already
noticed in Ref.  \cite{GillMassar}, for the maximum likelihood strategy
the optimal schemes can be surprisingly achieved by separable
measurements, and here we address this issue for covariant
measurements. Under the general assumption of square-summable
representation we derive a general "canonical form" for the optimal
measurements for pure input states, corresponding to a POVM which is
separable or entangled, depending on the group representation.

After introducing in Section \ref{s:theproblem} the precise
formulation of the covariant state estimation problem, in Section
\ref{s:ginte} we derive some useful mathematical identities for group
integrals which are then used to algebraically characterize covariant
measurements. This also helps us in deriving a simple upper bound for
the maximum likelihood in Section \ref{s:maxlik}, along with the
canonical form of the optimal measurement given in terms of the group
representation. Examples of the canonical form are given in Section
\ref{s:examp} in dimension $d<\infty$ for the group
$SU(d)$---corresponding to the estimation of an unknown pure
state---and in infinite dimensions for the estimation of displacements
on the phase space.  The case of multiple copies is then analyzed,
discussing the occurrence of entangled versus separable POVM's.  For
the estimation of displacements on the phase space, the case of two
copies experiencing opposite shifts in momentum is also analyzed---the
continuous-variables analogue of the measurement of direction of the
angular momentum with two antiparallel spins by Gisin and Popescu
\cite{gispop}. For coherent states it is shown that such a scheme
provides a better estimation of the displacement as compared to the
conventional case of identical displacements.
\section{The problem}\label{s:theproblem}
Whenever a quantum system $\System$ undergoes a physical
transformation belonging to a group $\gG$, its state is transformed
according to an appropriate representation of $\gG$ on the Hilbert
space $\sH$ of the system $\System$. In the following, we will
consider the case in which the group $\gG$ is a Lie group which acts
on $\sH$ by a (projective) unitary representation $\{U_g\}$, whereas
the initial state---also called \emph{seed} state---is a pure state
$|\Psi\rangle$. Notice that the correspondence between transformed
states and group elements is generally not injective, since the state
$|\Psi\rangle$ may have a nontrivial stability group, say $\gG_\Psi$
(we say that a group element $h$ belongs to the stability group
$\gG_\Psi$ of $|\Psi\rangle$ when
$U_h|\Psi\rangle=e^{i\phi_h}|\Psi\rangle$, with $\phi_h$ a real
phase).  In this way the transformed states are in one-to-one
correspondence with the cosets $gG_\Psi$: in other words the
group-orbit manifold (obviously invariant under the group
representation $\{U_g\}$) is identified with the coset space
$\Klm=\gG/\gG_\Psi$. We see that in principle from the output state
$U_g|\Psi\rangle$ it is possible to estimate the group element $g$ of
the transformation $U_g$ only if the stability group $\gG_\Psi$ of the
input state $|\Psi\rangle$ is trivial. Otherwise, we can estimate the
coset $\param\in\Klm$ which is in one-to-one correspondence with the
output state $|\Psi_\param\rangle=U_{g(\param)}|\Psi\rangle$,
$g(\param)$ labeling any element of $\gG$ in the coset $\param$. In
the following we will denote by $\param_0\equiv eG_\Phi$ the coset
containing the identity element $e$, and the seed state is relabeled
accordingly as $|\Psi_{\param_0}\rangle\equiv|\Psi\rangle$. This
notation makes explicit the isomorphism between the coset space $\Klm$
and the {\em homogeneous} manifold of states $|\Psi_\param\rangle$
$\param\in\Klm$, i.e. on which the group acts transitively through its
unitary representation as
$U_g|\Psi_\param\rangle\propto|\Psi_{g\param}\rangle$ (apart from a
phase factor). In this way, the estimation of the parameter
$\param\in\Klm$ becomes equivalent to a problem of {\em covariant
  state estimation}, and it was proved \cite{olevo} that the optimal
probability distribution $p(\param|\param_0)$ of estimating $\param$
for input state $|\Psi_{\param_0}\rangle$ satisfies the identity
$p(g\param|g\param_0)=p(\param|\param_0)$, namely the probability
distribution on the manifold $\Klm$ for an input state
$U_g|\Psi\rangle$ is equal to the probability distribution for input
state $|\Psi\rangle$ but with the manifold shifted by $g^{-1}$.  In
the following we will suppose for simplicity that the group $\gG$ is
unimodular (i.e. the left invariant measure $\d g$ on $\gG$ is also
right-invariant) and the stability subgroup is compact.  According to
a theorem by Holevo \cite{olevo}, for square-integrable
representations the covariant estimation is described by a POVM $M$ on
the probability space $\Klm$ of the general form
\begin{equation}
\d  M(\param) = U_{g(\param)} \,\Xi \,
U^{\dagger}_{g(\param)}\, \d \param\,,
\end{equation}
where $\d \param$ denotes the invariant measure on $\Klm$
induced by invariant measure $\d g$ on $\gG$ \cite{note},  
and the positive \emph{kernel} operator $\Xi$ belongs to the
commutant $\gG_\Psi'$ of the stability group (i.e. $[\Xi,U_h]=0 ]\;,\quad \forall
h\in \gG_\Psi$), and satisfies the completeness constraint
\begin{equation} \label{constr}
\int_{\Klm} \d \param\,  U_{g(\param)} \Xi U^{\dag}_{g(\param)}\equiv
\int_\gG \d g\,  U_g \Xi U^{\dag}_g=\Id\,.
\end{equation}
The fact that $\Xi\in \gG_\Psi'$  guarantees that the POVM does
not depend on the particular choice of $g(\param)$.
\section{Group integrals of operators}\label{s:ginte}
The completeness constraint in (\ref{constr}) becomes particularly
simple with some abstract considerations on group integrals. Since the
group $\gG$ is unimodular, its unitary square-summable representations
satisfy Schur's lemma for any (generally infinite dimensional)
representation space $\sH$  \cite{grossmann}, namely:
\begin{itemize}
\item[] For any couple $\{U^{\mu}_g\}$ and $\{U^{\nu}_g\}$
of irreducible components of the representation with invariant subspaces 
$\sH_{\mu},\sH_{\nu}\subseteq\sH$, respectively,
every operator $O_{\mu\nu}: \sH_{\nu} \to \sH_{\mu} $ 
satisfying the identity $U^{\mu}_g\, O_{\mu\nu} = O_{\mu\nu} U^{\nu}_g \quad
\forall g \in \gG$ must be of the form
\begin{equation*}
O_{\mu\nu} =  \left\{ \begin{array}{ll}
                         k\Ism_{\mu\nu}, \qquad &  \mbox{for }\mu\sim\nu\,,\\
                         0, & \mbox{otherwise,} 
          \end{array}
    \right.
\end{equation*}   
where $\sim$ denotes equivalence of irreducible components, $k$ is a constant, 
and $\Ism_{\mu\nu}:\sH_{\nu}\to\sH_{\mu}$ is the isomorphism 
mapping the two equivalent components, namely
$U^{\mu}_g=\Ism_{\mu\nu} U^{\nu}_g\Ism ^\dag_{\mu\nu} \quad \forall g \in \gG$
($\Ism_{\mu\mu}$ is the orthogonal projector onto the invariant
irreducible subspace $\sH_{\mu}$). 
\end{itemize}
A simple consequence of Schur's lemma is the Wedderburn
decomposition of operators $O$ such that $\Tr[I_{\mu\nu}O]<\infty$
$\forall\mu,\nu$  \cite{zhelobenko}
\begin{equation}
\int_{G} \d  g\, U_g O U^{\dagger}_g = \sum_{\mu} 
\sum_{\nu\sim \mu} a_{\mu\nu} \Ism_{\mu\nu}\,.
\label{wedd1}
\end{equation}
Taking the expectation values of both sides of Eq. (\ref{wedd1}) on an arbitrary
element $|e^{(\mu)}_n\rangle$ of an orthonormal basis $\{|e^{(\mu)}_m\rangle\}$
for $\sH_{\mu}$ one has
\begin{equation}
a_{\mu\mu}=\int_{G} \d  g\,
\Tr[(U^{\dagger}_g |e^{(\mu)}_n\rangle \langle e^{(\mu)}_n| U_g)
O],\quad\forall n\;.\label{star}
\end{equation}
Applying now the Wedderburn decomposition to the group average of
projectors $|e^{(\mu)}_n\rangle \langle e^{(\mu)}_n|$ and using invariance of
the subspace $\sH_\mu$, one obtains
\begin{equation}
\int_{G} \d  g\,U^{\dagger}_g |e^{(\mu)}_n\rangle \langle e^{(\mu)}_n|
U_g=b_{\mu\mu}\Ism_{\mu\mu}, 
\label{bmumu}
\end{equation}
where $b_{\mu\mu}$ is a constant to be evaluated. We then have
\begin{equation}
a_{\mu\mu}= b_{\mu\mu} \Tr[\Ism_{\mu\mu} O]\,,
\end{equation}
where $b_{\mu\mu}$ can be determined by taking the expectation value of
both sides of Eq. (\ref{bmumu}) on any normalized vector in $\sH_\mu$,
in particular on the vector $|e^{(\mu)}_n\rangle$, leading to
\begin{equation}
b_{\mu\mu}=\int_\gG \d  g |\langle e^{(\mu)}_n|U_g|e^{(\mu)}_n\rangle |^2\,.
\label{eq:mumu}
\end{equation}
On the other hand, if the representations ${\mu}$ and ${\nu}$ are
equivalent, there are two orthonormal basis $\{|e^{(\mu)}_n\rangle \}$ and
$\{|e^{(\nu)}_m\rangle \}$ for $\sH_{\mu}$ and
$\sH_{\nu}$, respectively, such that $\Ism_{{\mu}{\nu}}=\sum_{n}
|e^{(\mu)}_n\rangle \langle e^{(\nu)}_n|$. Now, taking the matrix element of
both sides of Eq. (\ref{wedd1}) between vectors $|e^{(\mu)}_n\rangle$ and
$|e^{(\nu)}_n\rangle$ one has
\begin{equation}
a_{\mu\nu}=\int_{G} \d  g\,
\Tr[U^{\dagger}_g|e^{(\nu)}_n\rangle\langle e^{(\mu)}_n|U_g O]\,,\label{dummy1}
\end{equation}
The invariance of both subspaces $\sH_\mu$ and
$\sH_\nu$ along with Schur's lemma give the identity
\begin{equation}
\int_{G} \d  g\,U^{\dagger}_g |e^{(\nu)}_n\rangle \langle e^{(\mu)}_n|
U_g=b_{\nu\mu}\Ism_{\nu\mu},\label{dummy2} 
\end{equation}
for suitable constant $b_{\nu \mu}$ to be determined. Upon substituting
the last equation into Eq. (\ref{dummy1}) gives 
\begin{equation}
a_{{\mu}{\nu}} = b_{{\nu}{\mu}} \Tr [\Ism_{{\nu}{\mu}} O]\,,
\end{equation}
and the constant $b_{\nu\mu}$ can be determined by taking the matrix element of
Eq. (\ref{dummy2}) between vectors  $|e^{(\nu)}_n\rangle$ and
$|e^{(\mu)}_n\rangle$, namely
\begin{equation}
b_{{\nu}{\mu}}= \int_\gG \d  g\,
\langle e^{(\nu)}_n|U_g^{\dagger}|e^{(\nu)}_n\rangle\langle e^{(\mu)}_n|U_g|e^{(\mu)}_n\rangle\,.
\end{equation}
Notice that for equivalent components $\mu\sim\nu$ for our choice of
bases one has
$\langle e^{(\mu)}_n|U_g|e^{(\mu)}_n\rangle=\langle e^{(\nu)}_n|U_g|e^{(\nu)}_n\rangle$,
whence $b_{\nu \mu}= b_{\mu\mu} = b_{\nu\nu} \equiv b_{\mu}$.
Summarizing, we have the decomposition
\begin{equation}
\begin{split}
\int_{G} \d  g\, U_g O U^{\dagger}_g\, &= \sum_{{\mu}}
b_{\mu} \sum_{{\nu} \sim {\mu}}\Tr[\Ism_{{\nu}{\mu}} O] \Ism_{{\mu}{\nu}}\,,\\
b_{\mu}&=\int_\gG \d  g |\langle e^{(\mu)}_n|U_g|e^{(\mu)}_n\rangle |^2\,.
\label{decomposition}
\end{split}
\end{equation}
\par If the group $\gG$ is compact and its measure $\d  g$ is
normalized (i.e. $\int_\gG \d  g\,=1$), then it is easy to show
that $b_{\mu} = \frac{1}{d_{\mu}}$, where $d_{\mu}=\dim(\sH_{\mu})$ 
(irreducible representations of compact groups are
finite-dimensional). In fact, summation over all $n$ in
Eqs. (\ref{star}) and (\ref{dummy1}) provide in a direct way the
values $a_{\mu\mu}=\Tr[\Ism _{\mu\mu}O]/d_\mu$ and 
$a_{\mu\nu}=\Tr[\Ism _{\mu\nu }O]/d_\mu $ 
for the coefficients in Eq. (\ref{wedd1}). On the other hand, the
derivation given above holds for unitary square-summable
representations, even with Dirac-orthogonal basis 
$\{|e^{(\mu)}_x\rangle \}$ for $H_\mu $, namely $\langle e^\mu _x 
|e^\mu _{x'} \rangle =\delta (x-x') $.  
The coefficients $b_{\mu}^{-1}$ are generally
non-integer, are called \emph{formal dimensions}, and
carry information about the structure of the irreducible components of
the group representation.
\section{Measurements with maximum likelihood}\label{s:maxlik}
We will now consider measurements which maximize the likelihood, namely
the conditional probability density $p(\param|\param)$ of having the
outcome equal to the true value for any $\param$. Because of
covariance this optimality criterion is equivalent to maximize the
\emph{likelihood functional} $\Lik_{\rho}[\Xi]= \Tr[\Xi
\rho]$ with $\rho = |\Psi \rangle \langle \Psi|$, $|\Psi \rangle$
being the input state.
\par Notice that the general solution to the maximum likelihood problem, which at first
sight may appear of limited value, is actually equivalent to the
solution of any quantum estimation problem with positive summable
"goal"-function $f(\estparam,\param)$ (the "goal"-function is the
opposite of the customary "cost"-function $-f(\estparam,\param)$ \cite{Helstrom76}).
This consists in associating to each measurement outcome $\estparam$ a "score"
$f(\estparam,\param)$, with the function $f(\estparam,\param)$
increasing versus $\estparam$ for  $\estparam$ approaching the true value
$\param$. Then, the optimal measurement is the one which maximizes the average score. 
In a covariant estimation problem  a meaningful
goal function must satisfy the invariance property
$f(\estparam,\param)=f(g \estparam,g \param)$ $\forall g\in \gG$, and this
allows to define a function $h(\estg,g)$ on the group via the relation $h(\estg,g)
\equiv f(\estg\param_0,g\param_0)$ for fixed $\param_0$. Then, the
function $h$ is positive (bounded from below), summable, and satisfies
$h(\estg,g)=h(g^{-1}\estg,e)$, $e$ denoting the identity
transformation. Now, thanks to covariance the average score can 
be written as 
\begin{equation*}
\begin{split}
\bar s =&\int_\gG \d  g\, h(g,e) \Tr [\rho U_g \Xi
U_g^{\dagger}]\\ =& \left[ \int_\gG \d  g\, h(g,e) \right]\,
\Lik_{\map{M}(\rho)} [\Xi]
\end{split}
\end{equation*}
where 
\begin{equation*}
\map{M}(\rho)=\frac{\int_\gG \d  g\,
h(g,e) U^\dagger _g \rho U_g}{\int_\gG \d  g\, h(g,e )}
\end{equation*}
is a completely positive trace preserving map. Therefore, the
maximization of a goal function can be viewed as a maximum likelihood
scheme on the transformed state $\map{M}(\rho)$, and depending on the
form of the function $h$ the choice of the input state may be
restricted to special states, possibly mixed. Nevertheless, in this paper we will give a
complete solution only for pure input states.
\par The problem is now
to find a positive operator $\Xi$ which maximizes the likelihood
functional $\Lik_{\rho}[\Xi]= \Tr [\Xi \rho]$, and, at the same
time, satisfies the completeness constraints (\ref{constr}).  Once
an optimal $\Xi$ is found, the presence of a nontrivial stability
group  $\gG_\Psi$ for $|\Psi\rangle$ can be taken into account by replacing
$\Xi$ with its group average over $\gG_\Psi$
\begin{equation}
\overline{\Xi}=\frac{\int_{G_\Psi} \d  g\,  U_g \Xi U^{\dagger}_g}{\int_{G_\Psi} \d  g}\,.
\end{equation}
Notice that the value of the likelihood functional remains unchanged
after this replacement, and the group average is still optimal (it is easy to show that the same occurs with
$\map{M}(\rho)$ in the case of a general goal function). As a
consequence of the Wedderburn decomposition (\ref{decomposition}), the
completeness constraint (\ref{constr}) for $\Xi$ can be written as
\begin{equation} \label{simpleconstr}
  \Tr[\Ism_{{\mu}{\nu}} \Xi] = \delta_{\mu\nu} b_{\mu}^{-1} \qquad
  \forall {\mu} \sim {\nu}\,.  
\end{equation} 
It is now convenient to decompose the input state $|\Psi\rangle$ over
the invariant subspaces $\sH_\mu$ of the representation as 
$|\Psi\rangle =\sum_{{\mu}} c_{\mu} |\Psi_{\mu}\rangle $. This allows
us to simply derive the following chain of inequalities
\begin{equation*}
\begin{split}
&\Lik_\Psi[\Xi]=\sum_{{\mu},{\nu}} c_{\mu}^* c_{\nu} \langle\Psi_{\mu}|\Xi|\Psi_{\nu}\rangle \leqslant \sum_{{\mu},{\nu}} |c_{\mu}|  |c_{\nu}| |\xi_{{\mu}{\nu}}| \\
&\leqslant \sum_{{\mu},{\nu}} |c_{\mu}|  |c_{\nu}| \sqrt{\xi_{{\mu}{\mu}} \xi_{{\nu}{\nu}}}
\leqslant \left( \sum_{\mu} |c_{\mu}| \sqrt{b_{\mu}^{-1}} \right)^2
\leqslant \sum_{\mu} b_{\mu}^{-1}\,,
\end{split}
\end{equation*}
where the sums range in the set $\set{M}_\Psi$ of all invariant
subspaces which are nonorthogonal to $|\Psi\rangle$, $\Lik_\Psi[\Xi]$
denotes the likelihood functional defined by the pure state
$|\Psi\rangle$, and $\xi_{{\mu}{\nu}}$ denotes the matrix element
$\langle\Psi_{\mu}|\Xi|\Psi_{\nu}\rangle $.  The first inequality can
be saturated by the choice
$\xi_{{\mu}{\nu}}=e^{i(\vartheta_{\mu}-\vartheta_{\nu})}
|\xi_{\mu\nu}|$ where $\vartheta_{\mu}$ is the phase of $c_{\mu}$.
The second inequality is a necessary condition for positivity of
$\Xi$, and saturates for
$|\xi_{\mu\nu}|=\sqrt{\xi_{\mu\mu}\xi_{\nu\nu}}$ (notice that this
inequality is not also a sufficient condition for positivity, whence
the positivity of the optimal $\Xi$ must be checked a posteriori). The
third inequality is due to the fact that $\xi_{\mu\mu} \leqslant \Tr
[\Ism_{\mu\mu} \Xi]=b_{\mu}^{-1}$.  Finally, the last Schwartz
inequality sets the following general upper bound for the maximum
likelihood of covariant measurements
\begin{equation}
\Lik_\Psi[\Xi]\leq\sum_{\mu\in\set{M}_\Psi}b_\mu^{-1}\,.
\label{eq:bound}
\end{equation}
In the case of a compact group
the inequality (\ref{eq:bound}) implies that the likelihood is
always less than the sum of dimensions of invariant subspaces
supporting $|\Psi\rangle$. For infinite dimensions, on the other hand,
the bound (\ref{eq:bound})  and the likelihood itself may diverge. One
can see now that the following choice of the operator $\Xi$
\begin{equation}\label{eta}
\Xi= |\eta\rangle \langle\eta|,\qquad|\eta\rangle =\sum_{\mu\in\set{M}_\Psi} e^{i\vartheta_{\mu}}
\sqrt{b_{\mu}^{-1}} |\Psi_{\mu}\rangle\,,
\end{equation}
attains the bound $(\sum_{\mu\in\set{M}_\Psi} |c_{\mu}|
\sqrt{b_{\mu}^{-1}})^2$ for the likelihood functional. Note that, if
$|\Psi\rangle $ has no component in some irreducible subspace
$\sH_{\nu}$, then the operator $\Xi$ must be extended to the
whole space $\sH$, in order to fulfill the constraints
$\Tr[\Ism_{\mu\mu}\Xi]=b^{-1}_{\mu}$ for all $\mu$. Obviously, such
extension is generally not unique, e.g. one can take
\begin{equation}
\Xi= |\eta\rangle \langle\eta| + \sum_{\nu\not\in\set{M}_\Psi}
b_{\nu}^{-1} |\Phi_{\nu}\rangle \langle\Phi_{\nu}|,\label{kernel-eg}
\end{equation}
where $|\Phi_{\nu}\rangle $ is any normalized vector in $\sH_{\nu}$,
which both guarantees $\Xi\ge 0$ and satisfies the constraints
$\Tr[\Ism_{\mu\mu} \Xi]=b^{-1}_{\mu}$ for all $\mu$. 
Notice that the presence of equivalent representations in
Eq. (\ref{kernel-eg}) generally improves the likelihood (this feature
was missed in Refs.  \cite{peres,tapia1,tapia2}).
\par If there are no equivalent representations in the decomposition of
$|\Psi\rangle $, then the kernel (\ref{kernel-eg}) averaged over the
stability subgroup $\gG_\Psi$ of $|\Psi\rangle $ is optimal. However, in
the presence of equivalent representations, one also wants the
off-diagonal constraints $\Tr[\Ism_{\mu\nu} \Xi]=0$ to be satisfied
$\forall \mu \sim \nu$. One can see that the kernel in
Eq. (\ref{kernel-eg}) satisfies also the off-diagonal constraints
when the decomposition $|\Psi\rangle =\sum_{{\mu}} c_{\mu}
|\Psi_{\mu}\rangle$ satisfies
\begin{equation}
\langle\Psi_{\mu}| \Ism_{\mu\nu} |\Psi_{\nu}\rangle = 0,\quad
\mu\sim\nu.\label{vanishingoff}
\end{equation}
As shown in the Appendix, the subspaces carrying equivalent
irreducible components of the representation can always be chosen in
such a way to satisfy Eq. (\ref{vanishingoff}). It is worth noticing
that the present "canonical" form for maximum likelihood measurements
generalizes the case of the optimal covariant phase estimation given
by Holevo  \cite{olevo}, further generalized in Ref.  \cite{DMS}.
Finally, notice that the result derived here also holds for discrete
groups, such as the permutation group or
$\mathbb{Z}_d\times\mathbb{Z}_d$ by just substituting integrals with
sums.
\section{Examples}\label{s:examp}
While it is obvious that averaging the result over a number $N>1$ of
equally prepared identical copies always improves the precision of
estimation---either classically or not---a legitimate question is
whether non-independent measurements on copies can be exploited to
further enhance the precision, compared to this conventional
independent measurement scheme. For the maximum-likelihood strategy,
when measurements are performed independently on each copy, the value
of the likelihood is linearly bounded as follows
\begin{equation}
\begin{split}
&\Lik_{\rm av}^{(N)}\equiv p\left(\frac{\sum_{i=1}^{N} \param_i}{N}=
\param|\param\right)\\&=\int\dots\int \d \param_1\cdots \d \param_N\,
p(\param_1|\param) \cdots p(\param_N|\param)\\ & \times \delta
\left(\frac{\sum_{i=1}^N \param_i}{N}-\param\right)\\ &\leqslant N
\max_{\param_N} \{p(\param_N|\param)\}\equiv N \Lik^{(1)}\,,\label{semi}
\end{split}
\end{equation}
where $\Lik^{(1)}$ is the maximum likelihood of a single-site
measurement.\par As we will show in the following, the optimal
measurement on $N$ copies of the same state can surpass the linear
bound for this {\em semi-classical} scheme involving independent
measurements. Moreover, if we relax the restriction of identical
preparation, corresponding to an input state of the form
$|\Psi\rangle=|\psi\rangle ^{\otimes N}$, it is also likely that a
preparation in different states can lead to further improvement in the
estimation of the group-transformation $U_g$ that occurred on the
input state, since the decomposition of the global state
$|\Psi\rangle$ may involve more invariant subspaces than just those
belonging to the symmetric space. Finally, the breach of the
semi-classical limit in Eq. (\ref{semi}) does not necessarily need
entangled POVM elements, and, actually, in one of the following
examples the linear limit is overcome by a separable POVM.
\subsection{Universal state estimation}
\subsubsection{$SU(d)$-covariant estimation: pure state estimation} 
The estimation of a pure state in a finite dimensional Hilbert space
$\sH$ can be regarded as a covariant estimation with respect to the
defining representation of the group $SU(d)$, where $d=\dim(\sH)$.
Indeed, the orbit of a given pure state contains all pure states of
$\sH$.  Clearly, the optimal kernel is $\Xi= d |\psi\rangle
\langle\psi|$, according to Refs. \cite{Helstrom76,olevo}, and
consequently the value of the maximum likelihood is $\Lik^{(1)}= d$.
\subsubsection{Pure state estimation with $N>1$ copies in the same
  state} This corresponds to the case of estimation of the group
element $g\in SU(d)$ in the reducible representation $U_g^{\otimes N}$
with initial state $|\Psi\rangle =|\psi\rangle^{\otimes N}$. There are
inequivalent components corresponding to the symmetric subspace
$(\sH^{\otimes N} )_+$, along with all other permutation invariant
subspaces. Since $|\Psi\rangle $ belongs to the symmetric subspace
$(\sH^{\otimes N} )_+$ only, the optimal $\Xi$ is not unique, e.g. we
can take $\Xi= d_{S} (|\psi\rangle \langle\psi|)^{\otimes N} +
\Id_\sW$, where $d_S={d+N-1 \choose d}$ is the dimension of
$(\sH^{\otimes N})_+$ and $\sW$ is the orthogonal complement of
$(\sH^{\otimes N})_+$.  In any case with $N$ copies we have maximum
likelihood $\Lik^{(N)}=\dim(\sH^{\otimes N})_+$, and for $N>2$ the
semi-classical limit $N d$ is breached. Notice that the POVM is not
separable, due to the presence of the orthogonal projector $\Id_\sW$.
\subsubsection{$SU(d)$ estimation with two copies in different states}
In this case $|\Psi\rangle =|\psi\rangle |\phi\rangle $ can be
decomposed as $\sqrt{\frac{(1+s^2)}{2}}\, |\Psi_+\rangle +
\sqrt{\frac{(1-s^2)}{2}}\, |\Psi_-\rangle $, where
$s=|\langle\psi|\phi\rangle |$ and
$|\Psi_{\pm}\rangle=\frac1{\sqrt{2(1\pm
s^2)}}(|\psi\rangle|\phi\rangle\pm |\phi\rangle |\psi\rangle)$.  Then
the optimal kernel $\Xi$ is proportional to the projector onto the
vector $|\eta\rangle =\sqrt{d_+} |\Psi_+\rangle + \sqrt{d_-}
|\Psi_-\rangle $ and the likelihood takes the value
$(\sqrt{\frac{d_+(1+s^2)}{2}} +
\sqrt{\frac{d_-(1-s^2)}{2}})^2\leqslant d^2$ (by Schwartz inequality).
It is easily seen that this bound can be attained by choosing
$s^2=\frac{1}{d}$. The optimal POVM is separable (the optimal kernel
is actually \emph{factorized} as $\Xi=d^2 |\psi\rangle \langle\psi|
\otimes |\phi\rangle \langle\phi|$). No further improvement can be
achieved, since a likelihood greater than $d^2$ is not compatible with
the completeness of the POVM (in fact $\Lik_\Psi[\Xi] \leqslant \Tr
[\Xi]= d^2$).
\subsection{Weyl-Heisenberg covariant estimation}
\subsubsection{Estimation of displacement on the phase space} This case
corresponds to consider the Weyl-Heisenberg irreducible representation
$\{D(z)\}$ of the translation group on the complex plane, $D(z)$
denoting the displacement operator $D(z)=e^{za^{\dagger}-z^*a}$ with
$[a,a^{\dagger}]=1$. Being non-compact, the representation space $\sH$
is infinite dimensional.  Physically $D(z)$ represents a joint shift
of position and momentum of a quantum harmonic oscillator, and the
covariant state estimation corresponds to a joint measurement of
position and momentum.  Here one has $b=\int_{\Cmplx} \frac{\d ^2
  z}{\pi}\, |\langle n| D(z)|n\rangle |^2$, where $|n\rangle $ denotes
an element of any orthonormal basis for $\sH$, which we can
conveniently take as the set of eigenstates of the number operator
$a^{\dagger}a$. Choosing $n=0$ one obtains $b=1$, whence the optimal
kernel for initial state $|\psi\rangle$ is $\Xi=|\psi\rangle
\langle\psi|$ and the maximum likelihood is $\Lik[\Xi]= 1$. Notice
that for $|\psi\rangle=|0\rangle$ we get the well-known coherent-state
POVM describing the heterodyne measurement  \cite{YuenSha,Bilk-poms}.
\subsubsection{Estimation of displacement with 
identical shifts on $N>1$ quantum oscillators} 
This case corresponds to the tensor representation $\{D^{\otimes
N}(z)\}$ of the Weyl-Heisenberg group. The irreducible representations
can be easily obtained by the linear change of modes represented by
the unitary transformation
\begin{equation*}
V= e^{\phi_N[a_1^{\dagger}(a_2+\dots +a_N)-a_1(a^{\dagger}_2+ 
\dots + a^{\dagger}_N)]}\,,
\end{equation*}
with $\phi =\frac{1}{\sqrt{N-1}}\arctan\sqrt{N-1}$ so that
$VD^{\otimes N}(z)V^{\dagger}=D(\sqrt{N}z) \otimes\Id^{\otimes
  (N-1)}$.  Then the irreducible subspaces are given by $\sH_n=
\{V^{\dagger} |\varphi\rangle\otimes
|\Phi_n\rangle,\;|\varphi\rangle\in\sH\}$, where $\{|\Phi_n\>\}$ is an
orthonormal basis for $\sH^{\otimes (N-1)}$. The formal dimension
coefficients are easily obtained as follows
\begin{equation*}
\begin{split}
  b_{n}=&\int_{\Cmplx} \frac{\d ^2 z}{\pi} |\<0|\langle\Phi_n| VD^{\otimes N}(z) V^{\dagger} |0\> |\Phi_n\rangle|^2\\
  =&\int_{\Cmplx} \frac{\d ^2 z}{\pi}
  |\<0|\langle\Phi_n| D(\sqrt{N} z) \otimes \Id^{\otimes (N-1)} |0\>|\Phi_n\rangle|^2\\
  =&\frac{1}{N}\int_{\Cmplx} \frac{\d ^2 z}{\pi} |\langle
  0|D(z)|0\rangle |^2=\frac{1}{N}\,.
\end{split}
\end{equation*}
Since the invariant subspaces carry all equivalent representations
---the isomorphism between two of them is $I_{mn}= V^{\dag} (I \otimes
|\Phi_m\>\<\Phi_n|) V$--- the problem of choosing a suitable
decomposition of the initial state $|\Psi\>$ in irreducible
representations arises. In the general case, one should apply the full
construction showed in Appendix, while a simpler solution is possible
for states of the form $|\Psi\>=|i_1\>|i_2\>\cdots|i_N\>$. In this
case, one has only to write $V|\Psi\>=\sum_{i_1,i_2, \dots, i_N}
c_{i_1 i_2 \dots i_N} |i_1\>|i_2\>\cdots |i_N\>$, and to define
$|\Phi_n\>= C_n^{-1} \sum_{i_2 \dots i_N} c_{n i_2 \dots i_N} |i_2\>
\cdots |i_N\>$, where $C_n =\sqrt{\left( \sum_{i_2, \dots, i_N} |c_{n
    i_2 \dots i_N}|^2 \right) }$, obtaining the desired decomposition
$|\Psi\>= \sum_n C_n V^{\dagger}|n\>|\Phi_n\>$ (notice that
$\<\Psi_m|\Psi_n\>= \delta_{mn}$ since they are eigenstates
corresponding to different eigenvalues of the number operator).  The
value of the likelihood is then $\Lik[\Xi] = N(\sum_n C_n)^2$.  

\par We now consider two special cases.

\par\noindent i) $N$ copies of vacuum state $|0\rangle $: this case
corresponds to the estimation of the complex shift $z$ on the set
$\{|z\rangle ^{\otimes N}\}$ of $N$ copies of a coherent state
$|z\rangle$.  Here, the vacuum state $|\Psi\rangle =|0\rangle
^{\otimes N}$ belongs just to one invariant subspace, since
$V|\Psi\rangle =|\Psi\rangle $.  The optimal kernel is not unique, and
is given by any completion of $\Xi = N (|0\rangle \langle 0|)^N$.  The
likelihood value is $N$, namely the "semi-classical" value. For $N=2$
it can be proved that one of the optimal POVM's corresponds to
averaging the outcomes of independent heterodyne measurements on two
copies, while another optimal one corresponds to the independent
measurement of the position $\frac12(a_1+a_1^\dag)$ and the momentum 
$\frac{1}{2i}(a_2-a_2^\dag)$, taking as the outcome $\alpha=x+iy$,
where $x$ and $y$ are the two separate outcomes.

\par\noindent ii) Two copies of a number state: $|\Psi\>= |n\>|n\>$
  with $n>0$.  The maximum value of the likelihood is $\Lik[\Xi]= 2
  \left( \sum_{k=0}^n \frac{1}{2^n n!} \binom{n}{k} \sqrt{(2k)!
  (2n-2k)!}  \right)^2$, and numerical calculation up to $n=1000$
  shows an almost linear behavior versus $n$, much better than the
  semi-classical value. In the case of two copies of a one-photon
  state $|\Psi\rangle =|1\rangle |1\rangle $. Decomposing the seed
  state we obtain $|\Psi\rangle
  =-\frac{\sqrt{2}}{2}(V^{\dagger}|20\rangle + V^{\dagger} |02\rangle
  )$: an example of optimal kernel is then $\Xi= 2\left[ 2\,(|1\rangle
  \langle 1|)^{\otimes 2} + \sum_{i\neq 0,2} V^{\dagger} |0i\rangle
  \langle 0i| V \right]$, achieving a likelihood equal to 4, which is
  yet twice the semi-classical value.
\subsubsection{Estimation of displacement on two copies, 
with identical shifts in position and opposite shifts in momentum}
This case corresponds to the representation $\{V(z) = D(z) \otimes
D(z^*)\}$, which is reducible, but does not possess any irreducible
proper component in $\sH^{\otimes 2}$, and thus is beyond the
hypotheses of our general results. In fact, the irreducible
representations are all inequivalent, and make a continuous set, each
component being supported by the Dirac-normalized eigenvectors 
\cite{rc} $\frac{1}{\sqrt{\pi}}
 |D(w)\> \doteq \frac{1}{\sqrt{\pi}} \sum_{m,n} \<m|D(w)|n\> \quad
 |m\>|n\>$ of the normal operator $W=a\otimes
I-I\otimes a^{\dagger}$ (the heterodyne photo-current
\cite{Bilk-poms,YuenSha,shwa}).  Upon expanding the operators
$V(z) =\exp(zW^\dag-z^*W)$ over the Dirac-orthonormal basis, one has
\begin{equation*}
\begin{split}
&\int_{\Cmplx} \frac{\d ^2 z}{\pi}\, V(z) O V^{\dagger}(z)=
\int_{\Cmplx} \frac{\d ^2 z}{\pi}
\int_{\Cmplx} \frac{\d ^2 w}{\pi}
\int_{\Cmplx} \frac{\d ^2 w'}{\pi} \,\times \\
&e^{z(w-w')^*-z^*(w-w')}|D(w)\rangle\langle D(w)|O|D(w')\rangle\langle D(w')|\\ =&
\int_{\Cmplx} \frac{\d ^2 w}{\pi} |D(w)\rangle\langle
D(w)|O|D(w)\rangle\langle D(w)|\;,
\end{split}
\end{equation*}
namely a continuous version of the Wedderburn decomposition still
holds
\begin{equation*}
\int_{\Cmplx} \frac{\d ^2 z}{\pi}\, V(z) O V^{\dagger}(z) = 
\int_{\Cmplx} \d ^2 w\, a_{w} P_{w}\,,
\end{equation*}
for any $O$ such that $\Tr[P_w O]<\infty$, with $P_w = |D(w)\rangle
\langle D(w)|$ and $a_w = {\pi }^{-1} 
\Tr[P_w O]$ (in a proper mathematical
setting the integral over $w$ in the last equation should be
interpreted as a direct integral). The maximum 
likelihood covariant measurement for state estimation among the set
generated by the seed $|\Psi\rangle \in \sH^{\otimes 2}$ is given by
the \emph{entangled} kernel $\Xi= |\eta\rangle \langle\eta|$, where
\begin{equation*}
|\eta\rangle  = \int_{\Cmplx} \frac{\d ^2 w}{\pi} 
e^{i \theta _w} |D(w)\rangle\,,
\end{equation*} 
which is the analogue of Eq. (\ref{eta}) for continuous spectrum (as
in that previous case, $\theta _w$ is the phase of $\langle
D(w)|\Psi\rangle $). It is worth noticing that for $|\Psi\rangle =
|0\rangle|0\rangle$ the problem corresponds to estimating the amplitude
$z$ of the set of coherent states $\{|z\rangle |z^*\rangle\}$, and the
value of the likelihood for the optimal measurement is $4$, namely
twice the likelihood for the amplitude estimation for identical states
$\{|z\rangle |z\rangle\}$. The probability distributions are indeed
Gaussian in both cases, but the variance in this case is half the
variance of the Gaussian for the states $|z\rangle|z\rangle$. The
fidelity of the estimate is $2/3$ for the states
$|z\rangle|z\rangle$, while it is $4/5$ for $|z\rangle|z^*\rangle$. The last
example can be regarded as the "continuous-variables" analogue of the
measurement of the direction of two antiparallel spins by Gisin and
Popescu \cite{gispop}.
\subsubsection{Estimation of displacement 
  on one part of a bipartite entangled system} We consider here the
representation $\{D(z) \otimes I\}$ acting on two optical modes. In
this case the invariant subspaces are $\sH_n=\{|\psi\> \otimes
|\varphi_n\>, |\psi\> \in \sH\}$, where $\{|\varphi_n\>\}$ is any
orthonormal basis in $\sH$, all of them supporting equivalent
representations, and the formal dimensions $b_n$ are all equal to 1.
If we take a twin beam $|\Psi\>= \sqrt{1-x^2} \sum_n x^n |n\>|n\>$ as
initial state, its decomposition is trivial, and $|n\>|n\>$ are
precisely the components $|\Psi_n\>$ on the irreducible subspaces.
Then the optimal POVM is given by $|\eta\>=\sum_n |n\>|n\>$, namely it
is the two mode heterodyne POVM \cite{YuenSha,shwa,rc}.
Correspondingly, the value of the likelihood is
$\Lik[\Xi]=\frac{1+x}{1-x}$, showing a strong enhancement by the
effect of entanglement in agreement with Ref. \cite{lopar}.
\section{Conclusions}
By group theoretic arguments we have derived the class of measurements
of covariant parameters which are optimal according to the maximum
likelihood criterion. The optimization problem has been completely
resolved for pure states under the simple hypotheses of unimodularity
of the group and measurable stability group. The general method has
been applied to the case of finite dimensional quantum state
estimation with many input copies, and, for infinite dimensions, to
the Weyl-Heisenberg covariant estimation, also giving a
continuous-variables analogue of the estimation of direction on two
antiparallel spins by Gisin and Popescu. The analysis allowed us to
compare the "semi-classical" statistical scheme of repeated identical
measurements on identically prepared copies with non-independent
measurement schemes, such as the LOCC and the entangled schemes.  We
have seen that for the maximum likelihood criterion, the possibility
of surpassing the semi-classical statistical efficiency with the
number of copies $N$ is essentially related to two factors: i) the way
in which the dimension of $N$-fold tensor product Hilbert space
increases versus $N$; ii) its decomposition in irreducible subspaces.
Moreover, the non separability of the optimal measurement---either in
its POVM or in the optimal states---is strictly related to the
structure of the group representation.  \par We conclude by mentioning
that the optimal covariant estimation for mixed input states is still
an open problem, and an explicit analytical optimization seems a very
difficult task. For the case of phase estimation the problem can be
analytically solved in the special instance of states which are
\emph{phase-pure} \cite{frecciadeltempo}.  For a general covariance
group-representation the concept of phase-pure state can be
generalized by choosing a vector $|\Psi_{\mu}\rangle$ for each
invariant subspace such that for every $\mu\sim\nu$ one has
$\langle\Psi_\mu|\Ism_{\mu\nu}|\Psi_\nu\rangle=0$. Then every state
$\rho$ satisfying $\Supp\{\rho\} \subset \Span\{ |\Psi_{\mu}\rangle
\}$ and $\langle\Psi_{\mu}|\rho|\Psi_{\nu}\rangle =
e^{i(\param_{\mu}-\param_{\nu})}
|\langle\Psi_{\mu}|\rho|\Psi_{\nu}\rangle |$ behaves as a pure state
in all respects, so that the upper bound and the canonical form of the
optimal POVM still hold. It is likely that a generalization of this
approach may extend the validity of the present solution of the
covariant estimation problem for a special class of mixed states.
\section{Appendix}
In this Appendix we show how to chose invariant subspaces $\sH _\mu$
of equivalent irreducible representations in order to satisfy Eq.
(\ref{vanishingoff}), namely $\langle\Psi_{\mu}| \Ism_{\mu\nu}
|\Psi_{\nu}\rangle =0$, for the decomposition $|\Psi\rangle
=\sum_{{\mu}} c_{\mu} |\Psi_{\mu}\rangle$, namely in such a way that
all invariant subspaces effectively behave as supporting inequivalent
irreducible representations. This choice of invariant subspaces
guarantees that every operator $\Xi = |\eta\rangle \langle\eta|$,
where $|\eta\rangle $ is any linear combination of
$|\Psi_{\nu}\rangle$, will satisfy the constraints $\Tr [\Xi
\Ism_{\mu\nu}]=0$ for any $\mu \sim \nu$. This method will allow to
extend the general treatment of the phase estimation problem given in
Ref.  \cite{DMS} to any square-summable group representation.  \par
Let's consider an irreducible component of a unitary representation of
$\gG$ with multiplicity $m\leq\infty$, and denote by $\sH^{(\omega)}$
the invariant subspace carrying all equivalent irreducible components,
and by $\sH^{(\omega)}= \bigoplus_{\mu=1}^m\sH_\mu$ a given choice of
invariant orthogonal subspaces $\sH_\mu$, each carrying an equivalent
irreducible representation. Since inequivalent irreducible components
already satisfy Eq. (\ref{vanishingoff}), we can just focus attention
on the component $|\Psi_\omega\rangle$ of $|\Psi\rangle$ on
$\sH^{(\omega)}$.  \par Let's denote by $\Ism_{\mu\nu}$ the
isomorphisms mapping $\sH_\nu$ into $\sH_\mu$, and satisfying
$[\Ism_{\mu\nu},U_g]=0$ $\forall g \in \gG$.  As already mentioned, it
is always possible to choose an orthonormal basis $\Base_\mu
=\{|e^{(\mu)}_n\rangle \}$ for each subspace $\sH_\mu$ in such a way
that $\Base_\mu= \Ism_{\mu\nu} \Base_\nu$ for any $\mu,\nu$, where
equality between bases is defined element-wise, i.e.
$|e^{(\mu)}_n\rangle = \Ism_{\mu\nu} |e^{(\nu)}_n\rangle $ for all
$n$. We have now the following simple lemma.  \par {\bf Lemma} (Choice
of the decomposition into equivalent components). For each unitary
matrix $\{V_{\mu\nu}\}\in\matrix_m$ the linear combinations
$\Base'_\mu = \sum_\nu V_{\mu\nu} \Base_\nu$ provide a new
decomposition $\sH^{(\omega)}= \bigoplus_{\mu=1}^m\sH'_\mu$ of
$\sH^{(\omega)}$ into subspaces supporting equivalent irreducible
components, where $\sH'_\mu \equiv \Span(\Base'_\mu)$.  \par {\bf
  Proof.} The subspaces $\sH'_\mu$ are orthogonal. In fact, upon
defining $\Base'_\mu = \{|f^{(\mu)}_l\rangle \}$, we obtain
\begin{equation*}
\begin{split}
  \langle f^{(\mu)}_l| f^{(\nu)}_n\rangle =& \sum_{\alpha,\beta}
  \langle
  e^{(\alpha)}_l| V^*_{\mu\alpha} V_{\nu\beta} |e^{(\beta )}_n\rangle \\
  =& \delta_{ln} \sum_\alpha (V_{\nu\alpha} V^{\dagger}_{\alpha\mu})=
  \delta_{ln} \delta_{\mu\nu}\,.
\end{split}
\end{equation*}
Moreover, each $\sH'_\mu$ carries a representation equivalent to that
of, say $\sH_1$. In fact, the operator $S_{i1} \equiv \sum_\nu
V_{\mu\nu} \Ism_{\nu1}$ is indeed an isomorphism between the subspaces
$\sH_1$ and $\sH'_\mu$, since it defines a one-to-one correspondence
between them via $\Base'_\mu = S_{\mu1} \Base_1$, and commutes with
$U_g$ for all $g \in \gG$, since each $\Ism_{\mu\nu}$ commutes. This
proves that the spaces $\{\sH'_\mu\}$ provide a new orthogonal
decomposition $\sH^{(\omega)}=\oplus_\mu\sH'_\mu$ into invariant
subspaces carrying equivalent components of the representation.

\par Now, let's consider the component $|\Psi_{\omega}\rangle $ of
$|\Psi\rangle$ on $\sH^{(\omega)}$, and write its decomposition
using the set of bases $\{\Base_\mu\}$ as follows
\begin{equation}\label{simple}
|\Psi_\omega\rangle  = \sum_{\mu n} \Psi^{\omega}_{\mu n} |e^{(\mu)}_n\rangle\,.
\end{equation} 
We want to construct a new decomposition $\sH'_\nu$ of
$\sH^{(\omega)}$ such that the components of $|\Psi_\omega\rangle $ on
invariant subspaces satisfy Eq. (\ref{vanishingoff}), namely they
behave as belonging to inequivalent representations.  This can be done
as follows. Define $\infty\geq d=\dim (\sH _\mu)$ and consider the
$m\times d$ matrix $\Psi^\omega=\{\Psi^\omega_{\mu n}\}$. According to
Eq.  (\ref{simple}), $\Psi^\omega$ is a Hilbert-Schmidt operator, and
hence we can write the singular value decomposition
\begin{equation}\label{svd}
\Psi^\omega= V ^T \Sigma U\,,
\end{equation}
where $\Sigma$ is a $m\times d$ matrix with all vanishing off-diagonal
elements, and $V$ and $U$ are $m\times m$ and $d\times d$ unitaries,
respectively. From Eq. (\ref{svd}) one obtains $\Psi^\omega_{\mu n}=
\sum_{\nu l} V_{\nu\mu} \sigma_\nu\delta_{\nu l} U_{l n}$, where the
sums run from $1$ to $r=\hbox{rank}(\Psi^\omega )\leq
\min(m,d)$. Equation (\ref{simple}) then rewrites
\begin{equation}\label{sing}
|\Psi_\omega\rangle  = \sum_{\nu l}\sigma_\nu \delta_{\nu l}
 |g^{(\nu)}_l\rangle\,, 
\end{equation}
where 
\begin{equation}
|g^{(\nu)}_l\rangle  
= \sum_{\mu n} V_{\nu\mu} \,U_{l n}  |e^{(\mu)}_n\rangle\,.
\end{equation}
The new bases $\Base'_\nu =\{|g^{(\nu)}_l\rangle \}$ provide a new
decomposition in invariant subspaces $\sH'_\nu$ supporting equivalent
components.  In fact, starting from the set $\{\Base_\mu \}$ the first
unitary transformation $U$ over each basis preserves the relations
$\Base_\mu =\Ism_{\mu\nu} \Base_\nu $, whereas the second
transformation $V$, according to the previous lemma, gives the new
decomposition $\sH^{(\omega)}= \bigoplus_{\mu=1}^m\sH'_\mu$, with
$\sH'_\mu \equiv \Span(\Base'_\mu)$.  The state
$|\Psi_{\omega}\rangle$ in Eq. (\ref{sing}) then satisfies
\begin{equation*}
\langle\Psi_{\mu}|
  \Ism_{\mu\nu} |\Psi_{\nu}\rangle =0\,,  \mbox{ for }{\mu \sim \nu}\;
\end{equation*}
since the spaces $\sH'_\nu=\Span\{|g_k^{(\nu)}\rangle\}$ have been
chosen such that each component of $|\Psi_{\omega}\rangle$ on
$\sH'_\nu$ is just proportional to a single element $|g_l
^{(\nu)}\rangle$ of the orthonormal basis with different $l$ for
different $\nu$.
\par  Notice that, if the group is
non-compact and there is an infinite number of equivalent irreducible
subspaces, the spectrum of singular values of $\Psi$ may be
continuous, and the sums in the above derivation must be replaced by
integrals, with some care in the generalization of definition and
theorems.
\subsection*{Acknowledgments} This work has been 
sponsored by INFM through the project PRA-2002-CLON, and by EC and
MIUR through the cosponsored ATESIT project IST-2000-29681 and
Cofinanziamento 2003.


\begin{thebibliography}{99}
\bibitem{Nielsen2000} I. L. Chuang and M. A. Nielsen, {\em Quantum
    Information and Quantum Computation} (Cambridge University Press,
  Cambridge, 2000).
\bibitem{Helstrom76}C. W. Helstrom, {\it Quantum detection and
    estimation theory} (Academic Press, New York, 1976).
\bibitem{buggec} R. Derka, V. Bu\u zek, and A. K. Ekert, Phys. Rev.
  Lett. {\bf 80}, 1571 (1998).
\bibitem{bem} D. Bru\ss , A. Ekert, C. Macchiavello, Phys. Rev. Lett.
  {\bf 81}, 2598 (1998).
\bibitem{joint_sub} G. M. D'Ariano, C. Macchiavello, and M. F. Sacchi,
  J. of Opt. B: Quant. and Semiclass. Opt. {\bf 3}, 44 (2001).
\bibitem{gisinrev} N. Gisin, G. Ribordy, W. Tittel, and H. Zbinden,
  Rev. Mod. Phys. {\bf 74}, 145 (2002).
\bibitem{single} G. M. D'Ariano and H. P. Yuen, Phys. Rev. Lett. {\bf
    76}, 2832 (1996).
\bibitem{WoottersZurek} W. K. Wootters, W. H. Zurek, Nature {\bf 299},
  802 (1982).
\bibitem{Yuen} H. P. Yuen, Phys.\ Lett.\ A{\bf 113}, 405 (1986).
\bibitem{univest} G. M. D'Ariano, P. Perinotti, and M. F. Sacchi,
  Europhys. Lett. {\bf 65}, 165 (2004).
\bibitem{Nai} M. A. Naimark, Iza. Akad. Nauk USSR, Ser. Mat. {\bf 4},
  277 (1940).
\bibitem{busc} F. Buscemi, G. M. D'Ariano, and M. F. Sacchi, Phys.
  Rev. A {\bf 68}, 042113 (2003).
\bibitem{YuenSha} H. P. Yuen and J. H. Shapiro, IEEE Trans. Inf.
  Theory, {\bf 24}, 657 (1978); {\bf 25}, 179 (1979); {\bf 26}, 78
  (1980).
\bibitem{Bilk-poms} G. M. D'Ariano, {\em Quantum Estimation Theory and
    Optical Detection}, in {\em Concepts and Advances in Quantum
    Optics and Spectroscopy of Solids}, ed. by T. Hakio\v{g}lu and A.
  S. Shumovsky (Kluwer Academic Publishers, Amsterdam 1997), p.
  139-174
\bibitem{direzionespin} G. M. D'Ariano, P. Lo Presti, and M. F.
  Sacchi, Phys. Lett. A {\bf 292}, 233 (2002).
\bibitem{peres} A. Peres and P. F. Scudo, Phys. Rev. Lett.  {\bf 86}, 4160 (2001).
\bibitem{tapia1} E. Bagan, M. Baig, and R. Mu\~noz-Tapia, Phys. Rev.
  Lett.  {\bf 87}, 257903 (2001).
\bibitem{tapia2} E. Bagan, M. Baig, and R. Mu\~noz-Tapia, Phys. Rev. A
  {\bf 64}, 022305 (2001).
\bibitem{sasschind} L. Susskind and J. Glogower, Physics {\bf 1}, 49
  (1964).
\bibitem{olevo} A. S. Holevo, {\it Probabilistic and statistical
    aspects of quantum theory} (North-Holland, Amsterdam, 1982).
\bibitem{noent} C. H. Bennett, D. P. DiVincenzo, C. A. Fuchs, T. Mor,
  E. Rains, P. W. Shor, J. A. Smolin, and W. K. Wootters,  
Phys. Rev. A {\bf 59}, 1070
  (1999).
\bibitem{Chefles} A. Chefles, Phys. Rev. A {\bf 64}, 062305 (2001).
\bibitem{lopar} G. M. D'Ariano, P. Lo Presti, and M. G. A. Paris,
  Phys. Rev. Lett. {\bf 87}, 270404 (2001).
\bibitem{olevoasymptotic} A. S. Holevo, quant-ph/0307225 (2003).
\bibitem{GillMassar} R. D. Gill and S. Massar, Phys. Rev. A {\bf 61},
  042312 (2002).
\bibitem{gispop}N. Gisin and S. Popescu, Phys. Rev. Lett. {\bf 83},
  432 (1999).
\bibitem{note} For compact $\gG_\Psi$, which is always our case (even
  for non compact $\gG$), then an invariant measure on $\Klm$ can be
  explicitly constructed using the map $g(\param)$  \cite{olevo}. The
  measure $\d \param$ is defined by the requirement that $\int_\Klm
  f(\param)\d \param= \int_\gG f(g \param_0)\d g$, for any function
  $f$ integrable on $\Klm$, where $\param_0$ is a fixed element in
  $\Klm$.  Compactness of $\gG_\Psi$ ensures normalizability of $\d
  \param$, and invariance follows from left invariance of $\d g$.
  Moreover, right invariance of $\d g$ implies that $\d \param$ is
  independent on the particular choice of $\param_0$.
\bibitem{grossmann} A. Grossmann, J. Morlet, and T. Pauli, J. Math.
  Phys. {\bf 26}, 2473 (1985).
\bibitem{zhelobenko} D.P. Zhelobenko, \emph{Compact Lie Groups and
    Their Representations} (American Mathematical Society, Providence,
  1973).
\bibitem{DMS}G. M. D'Ariano, C. Macchiavello, and M. F. Sacchi, Phys.
  Lett. A {\bf 248}, 103 (1998).
\bibitem{rc}G. M. D'Ariano and M. F. Sacchi, Phys. Rev. A {\bf 52},
  R4309 (1995).
\bibitem{shwa} J. H. Shapiro and S. S. Wagner, IEEE J. Quantum
  Electron.  QE {\bf 20}, 803 (1984).
\bibitem{frecciadeltempo}G. M. D'Ariano, C. Macchiavello, P.
  Perinotti, and M. F. Sacchi, Phys. Lett. A {\bf 268}, 241 (2000).
\end{thebibliography}
\end{document}